\documentclass[twocolumn,showpacs,pra]{revtex4}
\usepackage{makeidx}
\usepackage{graphicx}
\usepackage{dcolumn}
\usepackage{bm}
\usepackage{amsmath}
\usepackage{amsfonts}
\usepackage{amssymb}

\begin{document}

\title{Controllable Coupling between Flux Qubit \\and Nanomechanical Resonator
by Magnetic Field}
\author{Fei \surname{Xue}}
\author{Y. D. \surname{Wang}}
\author{C. P. \surname{Sun}}
\affiliation{Institute of Theoretical Physics, Chinese Academy of Sciences,
Beijing, 100080, China}
\author{H. Okamoto}
\author{H. Yamaguchi}
\author{K. Semba}
\affiliation{NTT Basic Research Laboratories, NTT Corporation,
Atsugi-shi, Kanagawa 243-0198, Japan}

\date{\today }

\begin{abstract}
We propose an active mechanism for coupling the quantized mode of a
nanomechanical resonator to the persistent current in the loop of
superconducting Josephson junction (or phase slip) flux qubit. This
coupling is independently controlled by an external coupling
magnetic field. The whole system forms a novel solid-state cavity
QED architecture in strong coupling limit. This architecture can be
used to demonstrate quantum optics phenomena and coherently
manipulate the qubit for quantum information processing. The
coupling mechanism is applicable for more generalized situations
where the superconducting Josephson junction system is a multi-level
system. We also address the practical issues concerning experimental
realization.
\end{abstract}

\maketitle

\section{Introduction}

In recent years, great advances in improving the coherence of
superconducting qubit have made it a promising candidate for the
physical realization of quantum information processing. Single qubit
Rabi oscillation and Ramsey fringe have been observed and two qubits
entanglement is also achieved. Meanwhile, as an artificial two-level
atom, superconducting qubit is adjustable (e.g. by flux, bias
voltage, etc.) and scalable. These features are favorable for
quantum state engineering. A number of protocols are proposed to
engineer the superconducting qubit to form a quantum network. Among
them, a very intriguing and successful example is the circuit QED
architecture~\cite{Blais2004}. By coupling the Cooper pair box
(charge qubit) to the quantized field of a coplanar superconducting
transmission line, a macroscopic solid-state analog of cavity QED is
realized on chip. Most recently, vacuum Rabi oscillations are
observed in a coupling system of 3-JJ flux qubit and LC circuit
\cite{Johansson2006}. Quantum optical phenomena in traditional
cavity QED can be demonstrated in this solid state composite system.
Furthermore, due to its special structure, it offers a number of
advantages, such as strong coupling and easy controllability. Thus,
some protocols that cannot be realized previously in the optical
cavity QED now becomes possible~\cite{Wang2004b,Wang2005}.

The circuit QED experiments motivate us to investigate the
possibility of substituting other quantum solid state devices for
the transmission line. It is much desirable to couple Josephson
junction qubit to a device with low energy consuming and small size.
If the strong coupling and easy controllability can also be
achieved, we get another favorable cavity QED structure. A possible
candidate of this solid state device is the nanomechanical resonator
(NAMR). Nanomechanical resonators of GHz oscillation have already
been observed. It is supposed that the nanomechanical resonator
enter the quantum regime at the attainable temperature of the
dilution refrigerator. The schemes of coupling Josephson charge
qubit or phase qubit to NAMR have already been proposed. Based on
these coupling mechanisms, several quantum state engineering
protocols are put forward
\cite{Wilson2004,Hopkins2003,Martin2004,Zhang2005,Zhou2006}.
However, due to the difficulty to reach quantum regime of NAMR,
those protocols have not been implemented experimentally yet. On the
other hand, the coupling mechanism of NAMR and flux qubit is also an
attractive problem since the flux qubit is supposed to have longer
coherence time as it is less affected by the charge fluctuation in
the structure. To our best knowledge, this has not been studied in
detail previously. Here, we present a novel mechanism of coupling
NAMR to flux qubit. As we present below, the coupling strength
between the NAMR and the flux qubit can be adjusted conveniently by
a coupling magnetic field and turned on and off within the coherence
time of flux qubit. Since the coupling magnetic field is independent
of the single qubit operation, it is possible to make it strong
enough even for GHz oscillation. Therefore our proposal is in
principle possible to approach the ``strong coupling regime" of
cavity QED at attainable temperature of dilution refrigerator. This
coupling system acts as an analog of cavity QED system with more
flexibility. We expect that it enables various applications to
quantum information processing and quantum state engineering.

This paper is organized as follows: in Sec.\ref{sec:coupling}, we
briefly review the setup of 3 Josephson-junction flux qubit and NAMR
as well as their experimental progresses. Then we get into the
coupling mechanism for flux qubit and NAMR. This coupling mechanism
can be equally applied to rfSQUID flux qubit and phase slip flux
qubit. In Sec.\ref{sec:spectrum}, the spectrum of the coupling
system is presented in the ``weak coupling" and the ``strong
coupling" limits respectively. The readout and quantum nondemolition
measurement for the flux qubit is studied in
Sec.\ref{sec:readout-QND}. We also consider the application of this
coupling mechanism in quantum computation in
Sec.\ref{sec:quant-computation} and generalize the coupling system
to beyond spin-boson model in Sec.\ref{sec:beyond-sb}. The possible
problems on experimental realization and their solutions are given
in Sec. \ref{sec:exp}. In Sec.\ref{sec:remarks}, some discussions
and remarks are included.

\section{coupling nanomechanical resonator and flux qubit by a transverse
magnetic field}

\label{sec:coupling}

\subsection{The 3-junction flux qubit}

Josephson charge qubit system has been used to couple with NAMR.
Here, we study another superconducting qubit system -- 3 Josephson
junction (3-JJ) flux qubit
\cite{Mooij1999,Orlando1999,Wal2000,Wal2003,Saito2004}. In contrast
with charge qubits, the flux qubit is far less sensitive to charge
fluctuations. Estimations show that the flux qubit have a relatively
high quality factor~\cite{Orlando1999,Wal2003}. The configuration of
flux qubit consists of a superconducting loop with three Josephson
junctions, and the Josephson coupling energy is much larger than the
charging energy for each junction. The quantum state of this system
is mainly determined by the phase degree of freedom. The Josephson
energy of the three Josephson junction loop reads
\begin{eqnarray}
U\left( \varphi _{1},\varphi _{2}\right) =&-& E_{J}\cos \varphi
_{1}-E_{J}\cos \varphi _{2}  \nonumber \\
&-&\alpha E_{J}\cos \left( 2\pi f-\varphi _{1}-\varphi _{2}\right) ,
\end{eqnarray}
where the constraint of fluxoid quantization has already been taken
into account. Here, $E_{J}$ is the Josephson coupling energy of two
identical junction and $\varphi _{1}$, $\varphi _{2}$ are phase
differences across the two junctions respectively. The Josephson
energy of the third junction is $\alpha E_{J}$, $f=\Phi _{f}/\Phi
_{0}$ with $\Phi _{f}$ the external flux applied in the loop and
$\Phi _{0}=h/2e$ the flux quantum. In the vicinity of $f=0.5$, if
$\alpha >0.5$, a double-well potential is formed within each $2\pi
\times 2\pi $ cell in the phase plane and the two lowest stable
classical states have persistent circulating currents
$I_{p}=2eE_{J}\sqrt{1-\left( 1/2\alpha \right) ^{2}}/\hbar $ with
opposite directions. Therefore, the flux qubit is also called
persistent current qubit. Within the qubit subspace spanned by
$\left\{ \left\vert 0\right\rangle ,\left\vert 1\right\rangle
\right\} $ ($\left\vert 0\right\rangle $ and $\left\vert
1\right\rangle $ denote clockwise and counterclockwise circulating
states respectively), the Hamiltonian of the qubit system reads as
\begin{eqnarray}
H_{f}=\omega _{f}\sigma _{z}+\Delta \sigma _{x}=\Omega \tilde{\sigma}_{z},
\end{eqnarray}
with $\omega _{f}=I_{p}\Phi _{0}\left( f-0.5\right) $ is the energy
spacing of the two classical stable states and $\Delta $ the
tunneling splitting between the two states, $\Omega =\sqrt{\omega
_{f}^{2}+\Delta ^{2}}$ and $\tilde{\sigma}_{z}=\cos \theta \sigma
_{z}+\sin \theta \sigma _{x}$, tan$\theta =\Delta /\omega _{f}$. The
offset of $f$ from $0.5$ determines the level splitting of the two
states and the barrier for quantum tunneling between the states
strongly depends on the value of $\alpha $. If the third junction is
replaced by a dc SQUID, both $f$ and $\alpha $ are tunable in
experiments by the applied flux or the microwave current
\cite{Mooij1999,Orlando1999}.

\subsection{The nanomechanical resonator}

The flexural modes of thin beams can be described by the so-called
Euler-Bernoulli equations~\cite{Clelandbook2002}. In our proposal
only the fundamental flexural mode of the NAMR is taken into
account. All the other modes have a much smaller coupling to flux
qubit and can be neglected~\cite{Armour2001,Armour2002}. In this
case, the NAMR is modeled as a harmonic oscillator with a high-Q
mode of frequency $\omega _{b}$. The Hamiltonian without dissipation
reads~\cite{Hopkins2003,Martin2004}
\begin{eqnarray}
H=\frac{p_z^{2}}{2m}+ \frac{1}{2}m\omega _{b}^{2}z^{2}.
\end{eqnarray}
In pursuing the quantum behavior of macro scale object the nano
scale mechanical resonator plays an important role. At sufficient
low temperature the zero-point fluctuation of nano mechanical
resonator will be comparable to its thermal Brownian motion. The
detection of zero-point fluctuation of the nano mechanical resonator
can give a direct test of the Heisenberg's uncertainty principle.
With a sensitivity up to 10 times the amplitude of the zero-point
fluctuation, LaHaye \textit{et al.} have experimentally detected the
vibrations of a 20-MHz mechanical beam of tens micrometers size
\cite{LaHaye2004}. For a 20-MHz mechanical resonator its temperature
must be cooled below 1 mK to suppress the thermal fluctuation. For a
GHz mechanical resonator a temperature of 50 mK is sufficient to
effectively freeze out its thermal fluctuation and let it enter
quantum regime. This temperature is already attainable in the
dilution refrigerator.

The lithographic technology for NAMR is rather mature. The important
advantages of NAMR are the potentially higher quality factor and
frequency comparable to superconducting qubit. Ever since the early
demonstration of a radio frequency mechanical resonator at Caltech
\cite{Cleland1996}, great advances have been made. The attainable
frequencies for the fundamental flexural modes can reach $590$ MHz
for the doubly-clamped SiC mechanical resonator of the size $1\times
0.05\times 0.05~\mu $m~\cite{Roukes2000} and $1$ GHz oscillation
frequency has also been measured~\cite{Huang2003}. It is argued that
quantized displacements of the mechanical resonator were observed
despite of some opposite opinions~\cite{Schwab2005}. For a $1$ $\mu
$m beam a quality factor $Q$ of $1700$ has been observed at a
frequency of $110$ MHz~\cite{Knobel2003}. In a carefully designed
antenna shape, Gaidarzhy \textit{et al.} have achieved $Q=11000$ for
$21$ MHz oscillation at the temperature of $60$ mK and $Q=150$ for
$1.49$ GHz oscillation at the temperature of $1$K with a
comparatively large double clamped beam~\cite{Gaidarzhy2005}. The
significantly small size ($\sim \mu $m) of the NAMR is also
favorable for incorporating it in the superconducting qubit circuit.

\subsection{The composite system with tunable coupling}

To achieve a ``strong" interaction, the coupling dynamical variable
usually should be the dominant one in the dynamics of the composite
system. For the Josephson phase qubit~\cite{Yu2002,Martinis2002},
the phase degree of freedom dominates the dynamics and the bias
current coupled with the phase is modified by the dilatational
motion of the piezoelectric dilatational resonator
\cite{Geller2005}. While for the Josephson charge qubit, the
coupling mechanism is that the resonator displacement modifies the
effective bias charge of a Cooper-pair box
\cite{Armour2002,Martin2004,Wang2004,Zhang2005}. These previous
investigations enlighten us to consider the coupling between
persistent current in superconducting flux qubit loop and the motion
of nano mechanical resonator.

Since the Josephson coupling energy of each junction in the flux
qubit is much larger than that in the charge qubit, the persistent
current in the loop could be about hundreds of nano ampere
\cite{Wal2000} in contrast with the critical current of the charge
qubit (usually about $20 \sim 50$ nA). The magnitude of this
persistent current naturally leads us to consider the magnetomotive
displacement actuation and sensing technique
\cite{Cleland1996,Clelandbook2002}. It is well known when a current
passes through a beam with conducting material, the perpendicular
arrangeed of an external magnetic field and the direction of the
current generates a Lorentz force in the plane of the beam. This is
just the actuation part of the magnetomotive technique. Meanwhile,
the resulted displacement of the beam under the Lorentz force
generates an electromotive force, or voltage, which serves as
measurement. Thus, if the doubly-clamped nano beam coated with
superconducting material is incorporated in the superconducting
qubit loop, the persistent current induces a Lorentz force with
opposite directions for clockwise and counterclockwise current. The
oscillation of the NAMR is modulated by these Lorentz force. In this
way, the quantized harmonic oscillation mode of the beam is coupled
to the quantum state of the flux qubit system. And this is just the
coupling mechanism considered in our paper.

\begin{figure}[tp]
\centering
\includegraphics[scale=0.7]{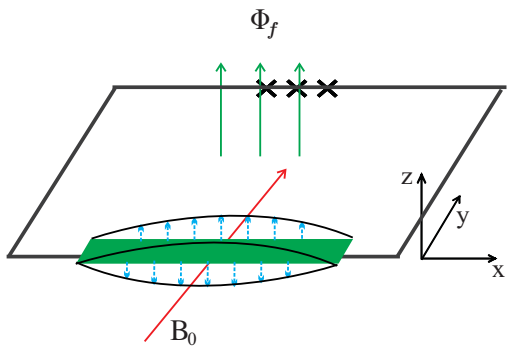}
\caption{(Color on line) A 3-Josephson-junction flux qubit loop is located
in the x-y plane and a NAMR is integrated in the loop(indicated by a green
line). The z-direction oscillation of the NAMR couples to the current in the
flux qubit loop by a transverse magnetic field $B_0$ in y-direction. Another
tunable magnetic flux $\Phi_f$ penetrates this loop tunes the free
Hamiltonian of the 3-JJ system. }
\label{fig:fig01}
\end{figure}

Our proposal is illustrated in Fig.\ref{fig:fig01}. A 3-JJ system is
fabricated on the x-y plane. The external applied magnetic flux
$\Phi _{f}$ is enclosed in the loop modulated by the control lines
(the lines are not plotted). The 4-JJ version of flux qubit system
can also be used here to allow the modulation of the effective
Josephson energy of the third junction and hence the tunneling
amplitude of the two current states. One side of the loop (indicated
by thick (green) rod) is suspended from the substrate and clamped at
both ends. This can be fabricated with a doubly-clamped
nanomechanical beam coated with superconductor or with the
superconductor itself as the mechanical resonator. A magnetic field
$B_{0}$ is applied in the $y$-direction. As we discussed above, the
circulating suppercurrent under the magnetic field generates a
Lorentz force in the z-direction. The magnitude of the force is
$B_{0}I_{p}L$, with $L$ the effective length of the resonator along
$x$-direction ($L=\xi L_{0}$ and $L_{0}$ is the actual length of the
resonator, $\xi $ a factor depending on the oscillation mode
\cite{Husain2003}, for the fundamental oscillation mode of a doubly
clamped beam $\xi \approx 0.8$). This force results a forced term in
the Hamiltonian, which reads $H_{fb}=Fz=B_{0}I_{p}Lz$. $\ $With the
two-level approximation of 3-JJ loop and the singlemode boson
approximation, the coupling is written as
\begin{eqnarray}
H_{fb}=g\left( a+a^{\dag }\right) \sigma _{z},
\end{eqnarray}
for $z \sim a + a^{\dagger}$. Here,
\begin{eqnarray}
g=B_{0}(t)I_{p}L\delta _{z}  \label{eq:g}
\end{eqnarray}
and $\delta _{z}=\sqrt{\hbar /2m\omega _{b}}$ is the amplitude of
zero point motion in z-direction of the NAMR, with $m$ the effective
mass of the resonator, $\omega _{b}$ the frequency of the
fundamental flexural mode; $a$ ($a^{\dag }$) is the creation
(annihilation) operator of the mode of the flexural motion in
z-direction. $\sigma _{z}=\left\vert 0\right\rangle \left\langle
0\right\vert -\left\vert 1\right\rangle \left\langle 1\right\vert $
is the Pauli matrix defined in the basis of $\left\{ \left\vert
0\right\rangle ,\left\vert 1\right\rangle \right\} $. We see that
this interaction $H_{fb}$ actually couples the two systems. Together
with the free Hamiltonian of flux qubit and NAMR, the Hamiltonian of
whole system reads
\begin{eqnarray}
H=\omega _{b}a^{\dag }a+\omega _{f}\sigma _{z}+\Delta \sigma _{x}+g\left(
a+a^{\dag }\right) \sigma _{z}.  \label{eq:Hamiltonian-original}
\end{eqnarray}

An important advantage of this coupling mechanism is the convenient
controllability. As seen from Eq.(\ref{eq:g}), the coupling constant
is directly dependent on the applied coupling magnetic field
$B_{0}$. Thus, both the magnitude and sign of the coupling constant
can be modified. What's more important is that the control parameter
$B_{0}$ in coupling coefficient Eq.(\ref{eq:g}) is independent from
the parameters of free Hamiltonian, such as bias voltage and
external magnetic flux $\Phi _{f}$. This means the free Hamiltonian
and the interaction Hamiltonian can be manipulated independently.
This full controllability is a rather favorable feature for quantum
state engineering and quantum information processing protocols. It
is in contrast with the coupling of charge qubit and NAMR, where the
coupling strength is controlled by the bias voltage which is also
the crucial parameter to determine the energy spacing of the charge
qubit. For example, for bang-bang cooling of NAMR by charge
qubit~\cite{Zhang2005}, the bias voltage should be set to certain
value to induce desirable damping. Therefore the on-and-off of the
interaction between the qubit and NAMR can only approximately
controlled by detuning and this can result in harmful reheating of
the NAMR. But in our present coupling mechanism, both the coupling
coefficient and the energy spacing are independent. Thus, this
``bang-bang" cooling protocol should be implemented more reliable by
the flux qubit and the NAMR with above coupling mechanism.

To estimate the coupling strength, we use the following parameters
in refs.~\cite{Lupascu2004,Martin2004,Aldridge2001}: $I_{p}=660$ nA,
$L_{0}=3.9$ $\mu $m, $\omega _{b}=100$ MHz, $\delta _{z}=2.6\times
10^{-13}$ m, $Q=2\times 10^{4}$ and assume the applied magnetic
field to be $B_0=5$ mT. Then we have $g\approx 4.01$ MHz. Hence we
see that the \textquotedblleft strong coupling" regime for cavity
QED is potentially realizable in our scheme. This regime requires
the period of the Rabi oscillation $1/g$, is much shorter than both
the decoherence time $1/\gamma $ of the two-level system and the
average lifetime $1/\kappa =Q/\omega _{b}$ of the \textquotedblleft
photon" in the \textquotedblleft cavity"~\cite{Haroche1992}. For
this composite system, the decoherence time for flux qubit is $1-10$
$\mu$s and the cavity lifetime is about $200$ $\mu $s, while the
Rabi oscillation time $0.016$ $\mu$s is much shorter than the two
lifetime scale. For GHz oscillation, the quality factor is rather
low\cite{Gaidarzhy2005} ($1.49$ GHz with $Q=150$). This
corresponding to a much shorter cavity lifetime (about $0.1$ $\mu$s.
However, the coupling strength can be increased by larger coupling
magnetic field. For example, if we take $B_0=50$ mT, the Rabi
oscillation period $1/g\approx 0.016$ $\mu$s which is still short
enough to reach ``strong coupling regime". Therefore, this protocol
might be promising in dilution refrigerator (several tens of
millikelvin).

\subsection{Phase slip flux qubit and NAMR}

\begin{figure}[bp]
\centering
\includegraphics[scale=0.7]{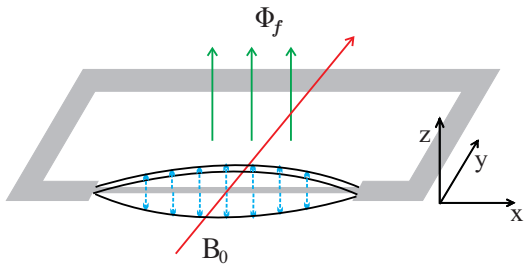}
\caption{(Color on line) A superconducting thin wire in the loop
acts as the center of phase slip flux qubit and a NAMR.}
\label{fig:phase slip}
\end{figure}

Most recently, a new type of flux qubit -- phase slip flux qubit is
proposed based on coherent quantum phase slip~\cite{Mooij2005,
Mooij2006}. Phase slip flux qubit is formed by a high-resistance
superconducting thin wire instead of the Josephson junctions. The
computational basis are also the two opposed persistent current
states. Our coupling scheme can be equally applicable to this type
of flux qubit. There are two important advantages to consider phase
slip flux qubit. First, since the superconducting thin wire can be
fabricated by a suspended carbon nanotube, it acts as the phase slip
center and the NAMR simultaneously. The circuit configuration is
simplified (see Fig. \ref{fig:phase slip}). Secondly, if one use
this qubit, one can free from any fluctuator due to imperfection or
two-level systems hidden in the dielectric layer of Josephson
junction.

\section{Energy spectrum of coupling system}

\label{sec:spectrum}

The Larmor frequency of superconducting qubit is about the order of
$10$ GHz, while the frequency of NAMR only reaches several hundred
MHz with quality factor $10^{4}$ at present stage. Thus, the
composite system of flux qubit and NAMR is in large detuning regime
of cavity QED, i.e., the following condition is satisfied
\begin{eqnarray}
\frac{g}{\left\vert \Omega-\omega _{b}\right\vert }\ll 1.
\end{eqnarray}
However, the superconducting flux qubit and the NAMR is
non-resonant, i.e. $\Omega \gg \omega _{b}$. This is in contrast
with Yale's circuit QED experiment where the Cooper pair box is
resonant with the 1D transmission line~\cite{Wallraff2004}. In the
following, we discuss the energy spectrum of our model in two
different regimes: $g\ll \omega _{b}$ (denoted as ``weak coupling")
and $g\approx \omega _{b}$ (denoted as ``strong coupling"). In our
proposal, the two regimes can be reached by varying the applied
coupling magnetic field $B_{0}$. And the energy spectrum are
qualitatively different from each other. It is notable that
Ref.\cite{Serban2006} has proved that the dispersive measurement
back action can be enhanced or reduced by cavity damping
respectively in the two regimes.

\subsection{``Weak coupling" and sideband spectrum}

\begin{figure}[tp]
\centering
\includegraphics[scale=0.8]{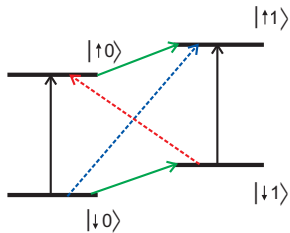}
\caption{(Color on line) The dressed energy level and transition diagram for
the weakly coupled system. }
\label{fig:fig02}
\end{figure}

For the parameters $B_{0}=5$ mT, $g=4.01$ MHz, both $g/\Omega $ and $g/\omega
_{b}$ are much smaller than $1$. In this case, the energy spectrum can be
calculated by Floquet approach or by Fr\"{o}lich transformation. After
performing a unitary transformation on the original Hamiltonian
(\ref{eq:Hamiltonian-original}), we get the effective Hamiltonian
\begin{eqnarray}
H_{\text{eff1}} &\approx &\omega _{b}a^{\dag }a+\Omega
\tilde{\sigma}_{z}+i \frac{g^{2}\sin 2\theta }{\omega _{b}}\left(
a^{2}-a^{\dag 2}\right) \tilde{
\sigma}_{y}  \nonumber \\
&&+\frac{g^{2}\sin \theta }{\Omega }\left( a+a^{\dag }\right) ^{2}\left(
\cos \theta \tilde{\sigma}_{x}+\sin \theta \tilde{\sigma}_{z}\right) .
\label{He1}
\end{eqnarray}
The spectrum are $n\omega _{b}+m\Omega $ plus some small
off-diagonal transition terms that are of order $O\left( {g}/{\omega
}\right) $ or $O\left( {g}/{\Omega }\right) $. As shown in
Fig.\ref{fig:fig02}, the energy levels of the two subsystems are
weakly perturbed by the coupling due to the large detuning. By
applying microwave pulse to induce the transition between those
levels, the blue sideband ($\left\vert 00\right\rangle \rightarrow
\left\vert 11\right\rangle $) and red sideband transition
($\left\vert 01\right\rangle \rightarrow \left\vert 10\right\rangle
$) can be observed in addition to the main zero-photon transition
$\left\vert 00\right\rangle \rightarrow \left\vert 10\right\rangle $
\cite{Goorden2004}. This atomic physics phenomenon has already been
observed in solid quantum system of flux qubit and dc SQUID
oscillator~\cite{Chiorescu2004}. For our proposal, similar spectra
are expected.

\subsection{``Strong coupling" and dispersive shift}

\begin{figure}[tp]
\centering
\includegraphics[bb=145 240 440 430,width=6cm,height=4cm,clip]{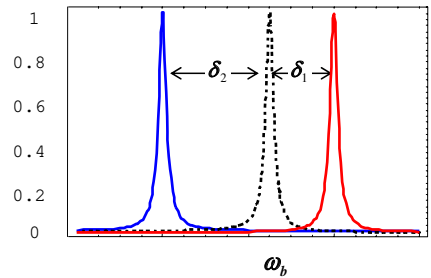}
\caption{Dispersive ``pull" of the frequency of NAMR with asymmetric
shifts from its original frequency. } \label{fig:fig04}
\end{figure}

With large magnetic field, for example, $B_{0}=100$ mT, then
$g=80.2$ MHz. The magnitude of coupling is comparable to the
characteristic energy scale of the NAMR. In this case, the coupling
term is not a perturbation with respect to the free Hamiltonian of
NAMR. Therefore, we can not use perturbation theory. However, since
the large detuning condition is still hold, we resort to adiabatic
elimination (or coarse-graining technique) to deal with this
problem. Since $\Omega \gg \omega _{b}$, the energy spectrum of the
whole system is roughly energy band structure. Then the spacing of
bands is determined primarily by $\Omega $ and the energy spacing
within each band is approximately $\omega _{b}$. By adiabatic
eliminating the transition between different bands, we can obtain
the effective Hamiltonian from eq.(\ref{eq:Hamiltonian-original})
\begin{eqnarray}
H=H_{e}\left\vert e\right\rangle \left\langle e\right\vert +H_{g}\left\vert
g\right\rangle \left\langle g\right\vert ,  \label{h1}
\end{eqnarray}
where the two component Hamiltonians are
\begin{eqnarray}
H_{e,g}=\omega _{e,g}A_{e,g}^{\dag }A_{e,g}\pm \Omega -g^{2}\cos ^{2}\theta
\frac{\omega _{b}}{\omega _{e,g}^{2}},
\end{eqnarray}
with the frequencies
\begin{eqnarray}
\omega _{e,g}^{2}=\omega _{b}^{2}\mp \frac{4g^{2}\omega _{b}\sin
^{2}\theta }{\Omega -\omega _{b}},
\end{eqnarray}
and the new bosonic operators $A_{e,g}$ are defined by
\begin{equation}
A_{e,g}=\mu _{e,g}a+\nu _{e,g}a^{\dag } + \eta_{e,g}.  \label{h4}
\end{equation}
In the above equation
\begin{subequations}
\begin{eqnarray}
\mu _{e,g} &=& \sqrt{\frac{\omega _{e,g}}{4\omega _{b}}}+\sqrt{\frac{\omega
_{b}}{4\omega _{e,g}}}, \\
\nu _{e,g} &=& \sqrt{\frac{\omega _{e,g}}{4\omega _{b}}}-\sqrt{\frac{\omega
_{b}}{4\omega _{e,g}}}, \\
\eta _{e} &=& \frac{g\cos \theta \sqrt{\omega _{b}\omega _{e}}}{\omega
_{e}^{2}}, \\
\eta _{g} &=& -\frac{g\cos \theta \sqrt{\omega _{b}\omega _{g}}}{\omega
_{g}^{2}}.
\end{eqnarray}
\end{subequations}
Thus, for different state of the flux qubit, the energy spectrum of
the NAMR is shifted by different value, or in other word, the flux
qubit pulls the cavity frequency by $\delta _{1}$ and $-\delta _{2}$
(see Fig.\ref{fig:fig04}) with
\begin{subequations}
\begin{eqnarray}
\delta _{1} &=&\sqrt{\omega _{b}^{2}+\frac{4g^{2}\omega _{b}\sin ^{2}\theta
}{\Omega -\omega _{b}}}-\omega _{b}, \\
\delta _{2} &=&\omega _{b}-\sqrt{\omega _{b}^{2}-\frac{4g^{2}\omega _{b}\sin
^{2}\theta }{\Omega -\omega _{b}}}.
\end{eqnarray}
\end{subequations}
In contrast with the dispersive limit discussed in circuit QED
\cite{Blais2004}, the dispersive shift here is asymmetric.

\section{QND measurement for flux qubit}

\label{sec:readout-QND}

Usually, the flux qubit is measured through a dc SQUID in the
underdamped regime that is inductively coupled
\cite{Wal2000,Saito2006} or directly coupled
\cite{Chiorescu2003,Bertet2005,Bertet2004}. The external applied
flux plus the flux induced by the persistent current decide the
switching current of the dc SQUID. By ramping a bias current to the
dc SQUID, the switching current can be recorded. When a continuous
microwave is resonant with the level spacing of the two eigenstates
of the flux qubit, the qubit is flipped and the switching current is
changed. This results in peak and dip in the switching current level
versus the applied flux. With this method, both the energy spectrum
and the dynamic evolution have been observed. During this
measurement process, the quantum information encoded in the qubit is
destroyed~\cite{Lupascu2004}. It would be favorable to design a
nondestructive and quantum nondemolition (QND)~\cite{Braginsky1992}
measurement protocol. A novel phase-sensitive microwave reflection
approach are now applied for the readout of superconducting qubits
\cite{Lupascu2004,Siddiqi2004,Siddiqi2006,Lupascu2006}. The
advantage of this method is that it directly probes the dynamics of
the Josephson plasma resonance in both the linear and nonlinear
regime without switching the detector Josephson circuit to
dissipative state. It succeeded to provide very fast and far less
destructive measurement of the qubit. However, the QND readout for
the flux qubit~\cite{Lupascu2006b} only works far away from the
optimal point, where the qubit coherence is destroyed very quickly.

Here, we indicate that our composite system can be used to perform QND
measurement on flux qubit. As we discussed in the previous section, in both
cases, the interaction between the NAMR and the flux qubit results in the mixed
energy spectrum for them. Therefore, through the spectroscopy measurement, the
quantum state of one system can be detected via the spectrum of the other one.
Especially, if the interaction Hamiltonian commutes with the free Hamiltonian
of the measured object and does not commute with that of the measuring device,
a QND measurement protocol can be implemented. This is just the case of
``strong coupling" limit in our proposal. In this case, by the spectroscopy of
the NAMR, the flux qubit state can be read out without perturbation. This can
be predicted from Eq.(\ref{h1}-\ref{h4}): the frequency of NAMR is $\omega
_{e}$ when the flux qubit is in the excited state, while its frequency is
$\omega _{g}$ when the flux qubit is in the ground state. The interaction
between the NAMR and the flux qubit commutes with the free Hamiltonian of the
flux qubit. Hence the measured probabilities of eigenstates are not perturbed
by this readout. Therefore, the spectroscopy measurement of NAMR provides a
high resolution QND measurement on the qubit state. It should be noticed that
this scheme does not work at the exact optimal point ($\sin\theta=0$ at the
optimal point). However, if only the operation point is a little bit shifted
away from the optimal point, e.g. $\omega_f\approx\Delta$, the resulting
frequency shift is observable (suppose $B_0=100$ mT, $\omega_b=100$ MHz and
$Q=2\times 10^{4}$, then $\delta_{1,2}\gg\kappa$) in the spectrum.

This frequency shift can be measured through the frequency
measurement of the NAMR. In principle, this can be done with
magnetomotive technique~\cite{Greywall1994,Gaidarzhy2005}. During
the measurement process, a perpendicular magnetic field and
oscillation current is applied on the NAMR. Then, the NAMR behaves
like a frequency dependent resistance. The largest effective
resistance is obtained when the NAMR is resonance with the
oscillation current. Thus the frequency of NAMR is inferred by the
resonance peak of its voltage between its two end when we vary the
frequency of the oscillation current in it. Another possible way to
the frequency measurement of NAMR is to use a single-electron
transistors (SET)~\cite {Blencowe1999,Zhang2001,Schwab2001}. The SET
does not require extra magnetic field which might induce unwanted
perturbation to the superconducting loop. It is supposed to have
very high sensitivity and is expected to reach the limit by the
uncertainty principle. For this method, the mechanical motion of the
NAMR couples to the SET through a lead on the NAMR that is close to
the island of SET. The motion of the NAMR modulates the coupling
capacitance between the lead and the SET. When there is a bias
voltage on the lead of NAMR the potential of the island near the
NAMR is modulated. The frequency of the mechanical motion is
detected through the conductance of the SET.

On the other hand, the measurement on high-frequency mechanical
oscillator has some practical difficulties as additional strong
coupled transducer is required to convert its dynamics to electronic
signals. There is an equally intriguing problem: to detect the
property of the NAMR via the measurement of superconducting qubit
since there have been some good measurement protocols of the latter.
For the coupling mechanism presented in this paper, the effective
Josephson energy is modified by the displacement of NAMR in a
similar way as the charge energy of SET being modified by the
displacement of NAMR. Thus, the flux qubit might be able to act as a
transducer to detect the state of NAMR. Most recently, a QND
measurement for NAMR via rfSQUID has been considered based on a
configuration similar to ours~\cite{Buks2006oct}.

\section{Applications in quantum computation}

\label{sec:quant-computation}

\begin{figure}[tp]
\centering
\includegraphics[scale=1]{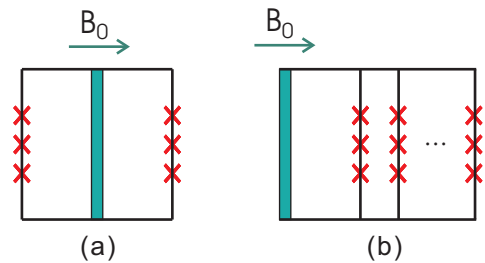}
\caption{(Color on line) The NAMR couples to (a) two or (b) multiple flux
qubits. The green narrow box represents the NAMR and the cross stands for
the Josephson junction. The coupling is controlled by a magnetic field
perpendicular to the NAMR in the coplanar of the NAMR and flux qubit.}
\label{fig:fig05}
\end{figure}

One of the possible applications of our proposal in quantum
computation is to couple two or more flux qubits together and to
realize two qubit logic gate. As shown in Fig.\ref{fig:fig05} (a),
the NAMR serves as a quantum data bus and the two identical qubit
loops are connected to it. The total Hamiltonian reads
\begin{eqnarray}
H=\omega _{b}a^{\dag }a+\sum_{i}\left( \omega _{f}\sigma _{zi}+\Delta \sigma
_{xi}+g\left( a+a^{\dag }\right) \sigma _{zi}\right),
\end{eqnarray}
where $\sigma _{zi}$ and $\sigma _{xi}$ are Pauli operators for the
$i$ th flux qubit. The coupling coefficient $g$ can be modulated by
the magnetic field $B_{0}$. If we fix $B_{0}$, the parameters for
manipulating single qubit (i.e., $\omega _{f}$ and $\Delta $)
operations are still tunable by adjusting $f$ and $\alpha$. This
offers a universal architecture to realize coherent two qubit
quantum logic gate~\cite{Makhlin2001}. Considering the inductance of
the loop $L_p$, we note that there is also direct magnetic coupling
induced by the sharing edge
$L_p(I_1+I_2)^2/2\sim\sigma_{z1}\sigma_{z2}$, here $I_1$, $I_2$ are
the currents in the 1st and 2nd qubit loop respectively. The order
of magnitude of this zz coupling is about $10$ MHz $\sim 100$ MHz.
At the degeneracy point, this always-on coupling is commute with the
whole Hamiltonian, therefore it is not very hard to deal with. Away
from the degeneracy point, the direct magnetic coupling may have
positive or negative effect with respect to concrete proposals.

More qubits can also be connected in the same way as shown in Fig.\ref
{fig:fig05} (b). By the tunable energy spacing of the flux qubit, we can
selectively couple two qubits through the NAMR. With this configuration, we can
realize the logic gate of two arbitrary flux qubits by adiabatically
elimination or by dynamically cancelation of the NAMR cavity mode
\cite{Raimond2001,Wang2004}.

\section{Beyond spin-boson model}

\label{sec:beyond-sb}

In the above discussion, we have described our coupling mechanism in
spin-boson regime where the 3-JJ superconducting loop is treated as
a quasi two-level system, i.e., a qubit. However, when the magnetic
flux $\Phi _{f}$ is tuned away from $\Phi _{0}/2$, the lowest two
energy levels cannot be isolated from other energy levels (see the
energy spectrum, for example, in ref.\cite{You2005}. Taking more
energy levels into consideration is advantageous to investigate many
intriguing phenomena that are traditionally studied in atomic
physics and quantum optics. For example, with the lowest three
energy levels, stimulated Raman adiabatic passage (STIRAP) can be
studied and some interesting behaviors, such as electromagnetic
induced transparency (EIT) and dark states, can be exhibited in the
coupling system. What's more, as the symmetry and the selection rule
of the 3-level superconducting loop are different from that of the
3-level natural atom, some novel features can be demonstrated, for
example, $\Delta $-type atom~\cite{Liu2005a} and persistent single
photon generation~\cite{Liu2006}.

\begin{figure}[tp]
\centering
\includegraphics[bb=10 10 280 230, width=7cm,clip]{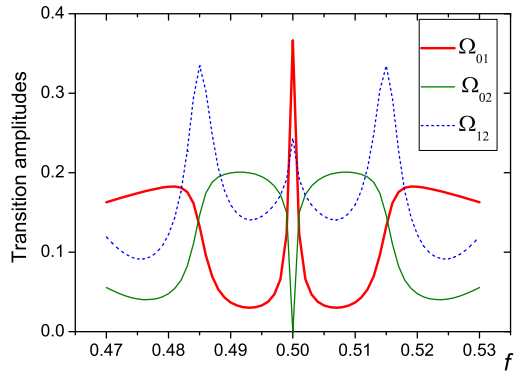}
\caption{(Color on line) The transition amplitudes between the lowest three
energy levels $\left\vert 0\right\rangle $, $\left\vert 1\right\rangle $ and
$\left\vert 2\right\rangle $ vary with the offset $f$.}
\label{fig:fig06}
\end{figure}

Our coupling protocol can also be generalized to the model of
multi-level atom in cavity. In this case, the dynamics of the 3-JJ
loop is not confined to the two-level subspace. With newly-defined
variables $\varphi _{p}=\left( \varphi _{1}+\varphi _{2}\right) /2$,
$\varphi _{m}=\left( \varphi _{1}-\varphi _{2}\right) /2$ and their
conjugate momentum $P_{p}$, $P_{m}$, the free Hamiltonian of the
3-JJ system has a similar form to that of a particle in a 2D
periodical potential~\cite{Orlando1999}:
\begin{eqnarray}
H_{f} &=&\frac{P_{p}^{2}}{2M_{p}}+\frac{P_{m}^{2}}{2M_{m}}+2E_{J}\left(
1-\cos \varphi _{p}\cos \varphi _{m}\right)  \nonumber \\
&&+\alpha E_{J}\left( 1-\cos \left( 2\pi f+2\varphi
_{m}\right) \right),
\end{eqnarray}
where $M_{p}=2C_{J}\left( \Phi _{0}/2\pi \right) ^{2}$ and
$M_{m}=M_{p}\left( 1+2\alpha \right) $, and $C_{J}$ is the
capacitance of the first and second junctions. Then the interaction
Hamiltonian is still $ H_{fb}=B_{0}ILz$ where $I$ is the current
flows through the NAMR~\cite {Liu2006}
\begin{equation}
I=\frac{2eC_{S}E_{J}}{\hbar C_{J}}\left( 2\cos \varphi _{p}\sin \varphi
_{m}-\sin \left( 2\pi f+2\varphi _{m}\right) \right) ,  \label{ip}
\end{equation}
with $1/C_{S}=\sum_{i}1/C_{i}$. The current $I$ induce the
transition between different eigenstates of the free Hamiltonian of
3-JJ system and numerical calculation can predict the transition
amplitudes. For example, if we only consider the lowest three energy
levels $\left\vert 0\right\rangle $, $\left\vert 1\right\rangle $
and $\left\vert 2\right\rangle $, the above Hamiltonian can be
written in a three dimensional subspace:
\begin{equation}
H=\omega _{b}a^{\dag }a+\sum_{i}\Omega _{i}\left\vert i\right\rangle
\left\langle i\right\vert +\lambda\sum_{i\neq j}\left( \Omega
_{ij}\left\vert i\right\rangle \left\langle j\right\vert
+h.c\right)(a+a^\dag)
\end{equation}
where $\vert i\rangle (i=0,1,2)$ is the $i$-th eigen level of the
3-JJ system and $\Omega _{i}$ is the corresponding eigenenergy
(usually we take $\Omega _{0}=0$). The coupling coefficient
$\lambda=B_0(t)L\delta_z$ and $\Omega _{ij}=\left\langle i\left\vert
I \right\vert j\right\rangle $. As shown in Fig.\ref{fig:fig06}, the
transition amplitudes $\Omega _{01}$, $\Omega _{02}$ and $\Omega
_{12}$ between the lowest three energy levels depend on $f$. That is
to say, we can control the transition between different energy
levels of the uncoupled system. Apparently, this feature can be used
in STIRAP technology and some other proposals based on three-level
atom with the quantized field~\cite{Bergmann1998}. For example, when
the superconducting loop is biased a little bit away from the the
optimal point, a $\lambda$ type three-level atom is formed
approximately. Following the same strategy of
ref.\cite{Siewert2005}, replacing the lowest three energy levels of
quantronium by the eigen-levels considered above, Fock-states of the
NAMR can also be generated.

\section{experimental considerations}
\label{sec:exp}

In the above discussion, we have assumed that the NAMR oscillates
only in the z-direction. However, to manipulate the 3-JJ flux qubit,
a bias magnetic flux $\Phi _{f}$ is applied through the loop in the
z-direction. This bias magnetic field also induces a Lorentz force
on the NAMR in y-direction. Therefore there exists an always-on
interaction between the flux qubit and the y-direction motion of the
NAMR $H_{fb}^{\prime }=g^{\prime }\left( a_{y}+a_{y}^{\dag }\right)
\sigma _{z}$, where $a_{y}$ denotes the annihilation operator of the
mode of the flexural motion in y-direction and the coupling strength
is
\begin{equation}
g^{\prime }=\Phi _{f}I_{p}L\delta _{y}/\Phi _{0},
\end{equation}
where $\delta _{y}$ is the zero point fluctuation of the NAMR in the
y-direction. However, Since
\begin{eqnarray}
\frac{g^{\prime }}{g}=\frac{B_{bias}}{B_{0}}\cdot \frac{\delta
_{y}}{\delta _{z}},
\end{eqnarray}
$B_{bias}=\Phi _{f}/S$, with $S$ the area enclosed by the loop of
3-JJ flux qubit, this additional coupling can be substantially
suppressed by a properly designed asymmetric structure of the NAMR
to set $\delta _{y}/\delta _{z}\ll 1$. This can be made when the
dimension of the NAMR in the y-direction is larger than that of the
z-direction. Then the dominant coupling is the one induced by the
magnetic field $B_{0}$. The bias field for the flux qubit at $\Phi
_{f}=0.5 \Phi _{0}$ is calculated as $19$ $\mu $T with the loop area
of Ref~\cite{Majer2005} or $250$ $\mu $T with a smaller loop of Ref
\cite{Chiorescu2003}. This corresponds a coupling strength of $12$
kHz or $160$ kHz (with $\delta _{y}=0.1\delta _{z}$), which is
always less than $10^{-1}$ of the coupling induced by the external
applied magnetic field $B_{0}$ even at the ``weak coupling"
situation discussed above. Therefore, this always-on coupling is
negligible for $B_{0}$ stronger than 1 mT.

Another possible difficulty in experiments might come from the
vibration of the sample (the substrate with the flux qubit and the
NAMR on it). If the controlling magnetic field $B_0$ in the
y-direction is generated by a coil located on another cold finger,
then in general, there may exist uncontrollable relative motion of
this sample cavity against the outside coil system. This relative
oscillation between the coil and the sample induces the fluctuation
of the controlling field. And more seriously, the torsional
oscillation of the sample will cause the deviation of the
controlling magnetic field $B_0$ away from y-direction, i.e., the
angle $\theta$ between $B_0$ and the plane of the qubit-NAMR loop
can not be zero strictly. Thus the none-zero component of $B_0$ in
z-direction $B_0 \sin\theta \sim B_0 \theta$ penetrating the loop
causes qubit energy fluctuations and decoherence. Roughly speaking,
this decoherence source is proportional to $B_0$. Therefore, one
need to optimize $B_0$ such that it is large enough to dominate the
coupling but not too large to induce strong decoherence from the
torsional vibration of the sample plane with respect to $B_0$.

\begin{figure}[tp]
\centering
\includegraphics[bb=84 268 404 575, width=5.1cm,clip]{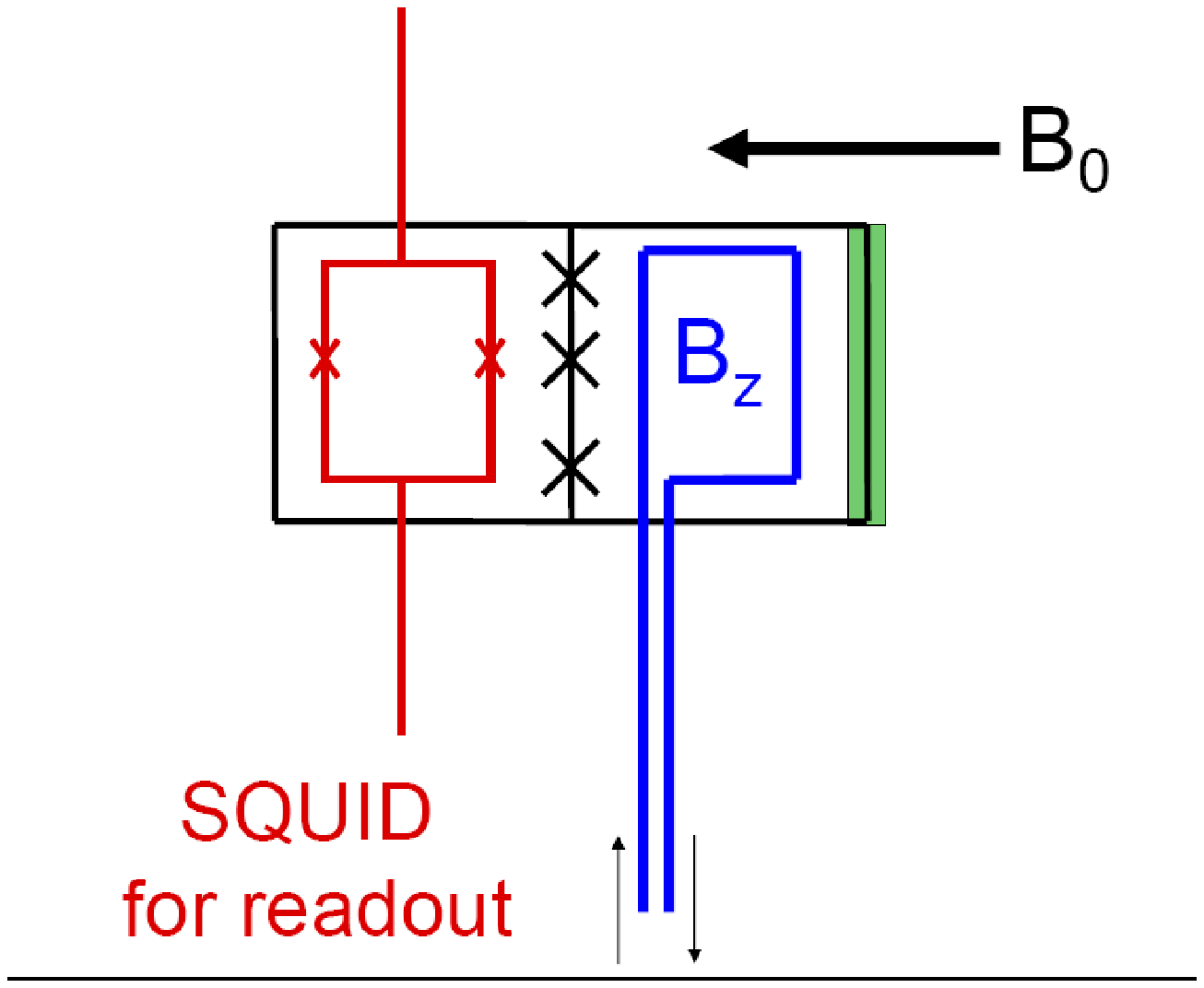}
\caption{(Color on line) The unexpected vibration induced
fluctuation in the bias magnetic field could be canceled by feeding
back the readout of the SQUID .} \label{fig:fig07}
\end{figure}

On the other hand, we could also reduce the fluctuation of the bias
magnetic field of flux qubit by improving the experimental setup.
For example, we could use an 8-shaped gradiometer qubit. By the
modified structure of the flux (as shown in fig.\ref{fig:fig07}), we
can cancel the uniform magnetic field fluctuations over the qubit in
the z-direction. The magnetic field threading the loop of the flux
qubit is measured by the readout SQUID. Then with suitable feedback
to compensate the fluctuation mentioned above, the bias magnetic
field in the qubit-NAMR loop can be stabilized. In this case, strong
magnetic field is attainable by the off-chip coil. However, scaling
up to many qubits is not so straightforward and needs further
consideration.

Another possible solution to overcome the above obstacles is to
prepare $B_0$ by a superconducting coil and fix it on the sample
chip at the dilution temperature. This method has advantageous that
the setup of the proposed controllable coupling mechanism need not
be modified and sufficient strong controlling magnetic field $B_0$
can be achieved.

\section{Discussions and remarks}

\label{sec:remarks}

In summary, we propose a novel solid-state cavity QED architecture
that can reach the ``strong coupling regime" based on a
superconducting flux qubit and an NAMR. In this composite system,
the quantized flexural mode of the NAMR is coupled with the
persistent current generated in the superconducting loop. The
coupling strength can be independently modulated by an external
magnetic field. We study the entangled energy spectrum of this
composite system and find that a QND measurement of flux qubit can
be made in the dispersive limit. This composite system can be scaled
up and the coupling mechanism can be extended to the case that the
superconducting junction is a multi-level system. We also carefully
examine the practical issues for experimental realization. The
controllable strong coupling and the scalability enable coherent
control over this system for the quantum information processing as
well as quantum state engineering. Besides, this cavity QED
architecture offers a new scenario to demonstrate the intriguing
phenomena of quantum optics in solid-state quantum device. It also
provides a possibility to test the quantization effect of mechanical
motion.

\acknowledgments
 This work is supported by the NSFC with grant Nos.
90203018, 10474104, 60433050 and 10574133. It is also funded by the
National Fundamental Research Program of China with Nos.
2001CB309310 and 2005CB724508. Y.D. Wang thank J. Q. You and P.
Zhang for helpful discussions. This work is also partly supported by
the JSPS-KAKENHI with grant Nos. 16206003 and 18201018. K. S. thanks
J. Plantenberg, C. J. P. M. Harmans, and J. E. Mooij for helpful
discussions.

\end{document}